# Infrared and Photoelectron Spectroscopy Study of Vapor Phase Deposited Poly (3-hexylthiophene)

*Haoyan Wei,[a),c] L. Scudiero,[b),c] Hergen Eilers [a),d]*

[a] Applied Sciences Laboratory, Institute for Shock Physics, Washington State University, Spokane, WA 99210, USA

[b] Chemistry Department and Materials Science Program, Washington State University, Pullman, WA 99163, USA

[c] These authors contributed equally.

**ABSTRACT**

Poly (3-hexylthiophene) (P3HT) was thermally evaporated and deposited in vacuum. Infrared spectroscopy was used to confirm that the thin films were indeed P3HT, and showed that *in-situ* thermal evaporation provides a viable route for contaminant-free surface/interface analysis of P3HT in an ultrahigh vacuum (UHV) environment. Ultraviolet photoelectron spectroscopy (UPS) as well as X-ray photoelectron spectroscopy (XPS) experiments were carried out to examine the frontier orbitals and core energy levels of P3HT thin films vapor deposited in UHV on clean polycrystalline silver (Ag) surfaces. UPS spectra enable the determination of the vacuum shift at the polymer/metal interface, the valence band maximum (VBM), and the energy of the π-band of the overlayer film. The P3HT vacuum level decreased in contrast to that of the underlying Ag as the film thickness increased. XPS and UPS data confirmed the chemical integrity (stoichiometry) of the polymer at high coverage, as well as the shift of the C 1s and S 2p binding energy peaks and the secondary-electron edge with increasing film thickness, indicating that band bending is present at the P3HT/Ag interface and that the measured onset of the valence band is about $0.8 \pm 0.05$ eV relative to the Fermi level.

## I. INTRODUCTION

Polythiophenes and their derivatives are an important class of conjugated polymers for a variety of applications in novel organic optoelectronics such as solid state electronics,[1,2] electroluminescent devices,[3,4] photovoltaics [5-7] and photodetectors.[8] In particular, poly (3-hexylthiophene) (P3HT), owing to its high drift mobility [9] (up to 0.1 cm$^2$ V$^{-1}$ s$^{-1}$ [1]), has been widely adopted as a hole transporting material in developing plastic electronics.[1,5] The device function and performance is largely dependent on the charge transfer process across the interfacial structure between organic semiconductors and electrodes. The adopted electrodes are usually metals (silver, gold and aluminum) and transparent conductive oxides (TCO) such as indium tin oxide (ITO). The charge transfer process is a function of the relative energy levels between the Fermi edge of the electrode and the HOMO and LUMO derived molecular orbitals in the valence band and conduction band regions, respectively.

Recently, nanoparticles of noble metals (silver, gold, etc) have been used in optoelectronic devices such as organic light emitting diodes (OLEDs) and solar cells. The metallic nanoparticles are either placed between two heterogeneous polymer layers [10] or incorporated into the bulk matrix.[11] The excitation of surface plasmons (collective oscillation of electron clouds), at the metal surface can dramatically improve the absorption and emission of light [12-14], and induce photoelectron injection.[15,16] In tandem solar cells, metal nanoclusters act as charge recombination centers connecting the front cell and the back cell in series.[17] The electronic structures of the vast interfaces play a key role in the optical and electrical performance of these nanocomposites.

Ultraviolet photoelectron spectroscopy (UPS) has been extensively used in studying various small organic molecules and polymeric materials. It is a powerful technique capable of depicting details of the valence electronic structures. In these measurements, organic semiconductors are usually deposited on top of a conductive substrate such as metals. To avoid surface/interface contaminations, the deposition of small organic molecules is usually done *in-situ* through vacuum evaporation. However, polymers are usually difficult to evaporate by vapor phase deposition techniques. Their large molecular weight and their long chains result in significant Van der Waals forces [18] and severe tangling between polymer chains. Instead, polymers such as P3HT are typically prepared from organic solvents through spin-coating onto underlying substrates.

UPS measurements of interfacial energy levels have been reported for solution-processed P3HT films.[19,20] However, this *ex-situ* method may introduce extra adventitious carbon or other contaminants at the interface and surface. In addition, P3HT is a *p*-type semiconductor and is very sensitive to ambient oxygen,[21] which strongly affects the electrical characteristics of semiconductors. These facts may have partially contributed to the wide range of values (4.3 eV to 5.2 eV) reported by different research groups [22-24] for the valence band maximum (VBM) relative to the vacuum level. In a recent effort to reduce the level of contamination, P3HT films were deposited *in-situ* by electrospray onto ITO and highly oriented pyrolytic graphite (HOPG).[25, 26] However, organic solvents were used in the dissolution of P3HT, and the potential adsorption of solvent molecules onto the substrates or their trapping within the deposited polymer films can not be overlooked.

The reversible thermochromic effect of P3HT from -60°C to melting state was demonstrated by Salaneck et al.[27] This implies that P3HT might be thermally evaporated in vacuum. In this contribution, we successfully demonstrate for the first time the evaporizability of P3HT using conventional thermal evaporation. Typically, vapor phase deposition of polymers undergoes a two-step process.[28,29] Upon thermal evaporation, polymer chains become fragmented as a result of pyrolysis. The re-polymerization occurs during their deposition on the substrate. Infrared (IR) spectroscopy measurements confirmed that the vapor phase deposited P3HT films largely retained their original chemical composition and structure. Using this technique, P3HT was deposited *in-situ* onto polycrystalline silver foil substrates in a vacuum chamber interfaced with a UPS/X-ray photoelectron spectroscope (XPS) analysis chamber, without breach of the vacuum environment and without the usage of organic solvents. High resolution XPS acquisitions of Ag 3d, C 1s and S 2p peaks and UPS valence band region of the P3HT/Ag system allowed the investigation of the chemical and electronic properties of P3HT.

## II. EXPERIMENTAL

A 300 nm thick P3HT film was deposited onto potassium bromide (KBr) substrates by vapor-phase deposition using an electron-beam heated effusion cell (Mantis Deposition Ltd.) in a high-vacuum chamber and used for infrared (IR) spectroscopy measurements. The base pressure of the vacuum system was $10^{-7}$ torr, and increased by two orders of magnitude during the deposition process. The P3HT evaporation process was very stable, comparable to the evaporation of metals except that much lower power was required. A schematic representation of the experimental setup is depicted elsewhere.[30] The nature of the chemical structure of the P3HT film was characterized by IR attenuated total reflection (ATR) on a Perkin Elmer Spectrum 2000 using a ZnSe crystal.

The P3HT films for UPS/XPS measurements were evaporated *in-situ* in a high-vacuum deposition chamber (base pressure of $10^{-8}$ torr) at a very low deposition rate (0.5-1 Å/min) from a Tantalum (Ta) boat purchased from R.D. Mathis Co. The deposition chamber is interfaced with the analysis chamber, ensuring sample deposition, transfer, and measurement all under ultrahigh vacuum without exposure to air. Initially, P3HT was kept below its melting point temperature of 230°C.[31] Subsequently it was heated to a temperature above its melting point and kept there for about 1 hr to remove potential impurities and other alien volatiles. A polycrystalline silver (Ag) foil obtained from Alfa Aesar was cut into squares of about 10x10 mm$^2$ and used as substrates. They were mechanically polished with a typical metal polish paste (Simichrome Polish), successively cleaned under sonication for several minutes in acetone and isopropanol alcohol, and finally rinsed with isopropanol alcohol and dried in air. The substrates were then immediately loaded into the XPS/UPS system and underwent Ar ion sputtering cleaning for at least 15 min.

XPS and UPS measurements were performed on a Kratos AXIS-165 multi-technique electron spectrometer system with a base pressure of $5 \times 10^{-10}$ torr. Achromatic X-ray radiation of 1253.6 eV (MgKα) was utilized as the XPS excitation source for acquiring all XPS photoelectron spectra. The binding energies were calibrated against the Au $4f_{7/2}$ peak taken to be located at 84.19 eV and Ag $3d_{5/2}$ peak at 368.46 eV.

The P3HT layer thickness was determined from the attenuation of the intensity of Ag 3d $_{5/2}$ peaks by using the ratio of the sample signal to that of a pure Ag substrate signal with $t = -\lambda E_{substrate} \cos\theta \ln(I_{sample}/I^0_{substrate})$ where $\lambda$ is the mean free path of the substrate material electrons of kinetic energy ($E_{substrate}$) through the sample overlayer and $\theta$ is the take-off angle, and by a quartz crystal microbalance. A value of $\lambda = 1.1$ nm was taken for Ag to be the total mean free path as calculated by Zemek.[32]

UPS data were collected with a homemade He lamp source which produces a resonance line He I (21.21 eV) by cold cathode capillary discharge. The spectra were acquired using a hybrid lens that focused the ejected electrons into the Kratos spectrometer. A bias of -20V was applied to the sample to shift the spectra out of the nonlinear region of the analyzer (0~10 eV kinetic energy). The energy resolution was determined at the Fermi edge of the Ag foil to be better than 150 meV.

XPS and UPS characterization of the samples after each growth step was performed by transferring the Ag substrate from the deposition chamber to the AXIS-165 via a series of UHV gate valves. After each growth cycle Ag 3d, C 1s and S 2p photoemission spectra were recorded. UPS was used to investigate the valence band region between 0 – 4.5 eV and to scan the full photoemission spectra to determine the magnitude of the induced dipole at the interface of Ag/P3HT layered structures.

## III. RESULTS AND DISCUSSION

Fig. 1 shows the infrared spectrum of a vapor phase deposited P3HT film on KBr substrate in the region from 4000 to 500 cm$^{-1}$. The spectrum is consistent with that of the powder P3HT material,[33,34] with a one-to-one correspondence in absorption peak positions, although there are minor differences in the intensity of some IR peaks. Three IR absorption regions at 3054, 1510-1430 and 832 cm$^{-1}$ were associated with vibrations of P3HT thiophene rings. The peak at 3054 cm$^{-1}$ is assigned to the aromatic C-H stretching vibration of the thiophene ring, while the C-H out-of-plane vibration of a 2,3,5-trisubstituted ring is located at 832 cm$^{-1}$. The two peaks in the band between 1510 and 1430 cm$^{-1}$ are assigned to the stretching vibration of the thiophene ring. Specifically, the absorption peak at 1456 cm$^{-1}$ is associated with a symmetric C=C ring stretching vibration while the peak at 1508 cm$^{-1}$ is related to an asymmetric C=C ring stretching vibration.[34] The intensity of the absorption peak at 1508 cm$^{-1}$ appears reduced in comparison to powder and solution processed materials. This implies that there might be preferential orientation of the polymer chains or some variations of the conjugation length.[35] The bands at 1510-1430 and 832 cm$^{-1}$ are characteristic of the 2,3,5-trisubstituted thiophene ring.[33]

The hexyl side chains are confirmed by the observation of three bands at 3000-2850 cm$^{-1}$, 1378 cm$^{-1}$ and 728 cm$^{-1}$. The peak at 1378 cm$^{-1}$ is assigned to the deformation vibration of terminal methyl groups -CH$_3$, while the peak at 728 cm$^{-1}$ confirms the rocking vibration of hexyl substituent methylene groups -(CH$_2$)$_5$-.[36,37] Three peaks within the band of 3000-2850 cm$^{-1}$ are the characteristics of C-H bonds on the aliphatic side chain, which have been assigned respectively to the asymmetric C-H stretching vibrations of -CH$_3$ (2955 cm$^{-1}$) and -CH$_2$- (2926 cm$^{-1}$) moieties, as well as the symmetric C-H stretching vibration in -CH$_2$- (2856 cm$^{-1}$) moieties.[38] The results above indicate that P3HT molecules underwent no or little change in composition and chemical structure, and their conjugated nature was largely retained upon thermal evaporation.

Fig. 2 depicts the thickness-dependent XPS spectra of C 1s and S 2p deposited on Ag foil. The spectra were normalized with respect to the highest peak and shifted vertically for clarity. Both C 1s and S 2p core level positions shift toward higher binding energy (BE) as the P3HT layer thickness increases. A maximum shift of 0.31 eV and 0.21 eV was measured respectively for C 1s and S 2p at coverage of 3.1 nm. This up-shift is most likely related to band bending at the interface of Ag/P3HT due to charge transfer to allow Fermi level alignment on both sides (see UPS discussion below). At very low coverage the S 2p spectra display two peaks, one at 164.5 eV and the second at 162 eV as determined by curve fitting. The peak at 164.5 eV is attributed to the S-C bonding in P3HT. The lower BE peak at 162 eV, similar to metal sulfides,[32] indicates the interaction between S and Ag.[39] As the polymer film thickness increases the lower BE peak associated with S-Ag bonding fades gradually and the higher BE peak, associated with S-C bonding, becomes more intense appearing as a doublet consistent with spin-orbit splitting in the expected ratio of 1:2 for pristine P3HT. The photoemission spectra of C 1s show a single peak that increases in intensity and shifts toward higher BE with increasing P3HT coverage. The absence of shake-up peaks on the high BE side of the main C 1s line is a clear indication that the conjugated π system in the P3HT solid film on Ag is not broken into smaller conjugation lengths.[27] At coverage of 3.1 nm, the measured C/S ratio (*ca.* 10.1:1) is in good agreement with the stoichiometric C/S ratio (10:1) of P3HT molecules. This finding further supports the FTIR conclusions that P3HT retained its chemical composition upon thermal evaporation.

Fig. 3 displays He I UPS spectra obtained for P3HT on polycrystalline Ag at each growth step. As P3HT films grow thicker, the Ag Fermi edge and the d-band features are gradually veiled by P3HT. Similar to previous spin-coated samples, only a broad band was observed for the vacuum deposited P3HT films between 5-12 eV which is associated with the σ states contributed from main backbones and hexyl side chains.[40,41] The secondary-electron edge at the high BE end and the onset of the valence band of P3HT position was directly determined from the He I spectra. The secondary cutoff energy allows for the estimation of the interface dipole (eD). By measuring the width of the photoemission spectrum for clean polycrystalline Ag from the secondary cutoff to the Fermi edge, the silver work function is determined to be 4.27 eV ($\Phi_{Ag}$) in good agreement with the literature value of 4.29 eV.[42,43] Using the same method, a work function of 3.71 eV ($\Phi_{P3HT}$) is obtained for the highest polymer coverage (3.1 nm). Thus their difference gives the maximum vacuum level shift of 0.56 eV ($\Phi_{Ag} - \Phi_{P3HT}^{3.1}$), which is the sum of the interface dipole (eD) and the band bending energy ($V_b$). A shift in the vacuum level is measured for coverage as low as 0.25 nm. By measuring this energy shift for the first deposition step the interface dipole is estimated to be on the order of 0.36 eV ($\Phi_{Ag} - \Phi_{P3HT}^{0.25}$). The band bending energy is thus obtained by subtracting this value from the 0.56 eV and gives a value of 0.20 eV which is in good agreement with the shift measured for S 2p (0.21 eV). Both interface dipole and band bending energies are higher than those obtained on HOPG.[26] The larger interface dipole is partly due to the stronger chemical interaction between P3HT and silver as evidenced in XPS measurements (Fig. 2). Finally, the onset of the valence band is determined to be 0.8 ± 0.05 eV for coverage of 0.44 nm at which vacuum shift starts to plateau (see right graph Fig. 3). This value is roughly twice the value of 0.41 eV obtained by Cascio et al. on HOPG.[26] Therefore, the valence band maximum (VBM) energy level of P3HT, referenced to the vacuum level, was determined to be 4.51 eV by adding work function $\Phi_{P3HT}^{3.1}$ and the valence band cutoff binding energy. Correspondingly, a conduction band minimum (CBM) of 2.81 ~ 2.61 eV is estimated by subtracting the optical bandgap $E_g^{opt}$ (*ca*. 1.7~1.9 eV as measured by UV-VIS spectroscopy,[23,44]) from VBM.

The right graph in Fig. 3 displays the valence band energy region of 0 - 4.5 eV for each deposition step and the clean polycrystalline Ag foil. The observed band features (at about 2 eV and 4 eV) are consistent with that of solution processed films.[26,27,45] The band, initially located at a binding energy of 3.4 eV (localized π-band), shifts toward higher BE as the polymer layer reaches the maximum thickness of 3.1 nm. At this coverage the band energy is *ca*. 3.95 eV, for a total shift of 0.55 eV. This localized π state constitutes S 3p and C 2p atomic orbitals.[40] The shift saturates at about 2 nm indicating that the P3HT layer has reached the characteristics of the bulk-like polymer material (as seen in Fig. 3). In addition, a weaker band appears in the region of 2-2.5 eV, which seems to remain at the same energy. This feature was also observed on HOPG and assigned to delocalized π states [27,46] contributed from C 2p and partial S 3p atomic orbitals.[40] An energy shift of the band at about 3.4 eV was also observed by Hao et al.[41] for spin-coated P3HT on gold. Several causes could be responsible for the up-shift of the localized π-band position: i)

The growth of the polymer forms clusters even at the monolayer coverage which will grant the P3HT the bulk like properties. ii) Polarization of the polymer layer due to a slightly different arrangement of the molecules as a result of the different geometries at the surface near the interface and the bulk.[47,48] This charge redistribution with layer thickness will manifest itself as energy shift in the band position. iii) The chemical interaction between deposited polymers and silver evidenced by foregoing XPS measurements. Further studies are needed here to understand this band energy shift. For instance to answer bullet i) topographic imaging investigation could be performed by setting up UHV atomic force microscopy on this system.

Fig. 4 schematically summarizes the correlative energy levels at the interface of P3HT/Ag. A large drop in vacuum level of P3HT is observed for the first three deposition steps (0.25, 0.32 and 0.44 nm) followed by a plateau (Fig. 3). Lowering of the vacuum level has been seen in most of the organic compounds thermally deposited onto a metallic substrate. The decrease is explained as a charge (electron) transfer from the material with the smaller work function to the material of larger work function (In this case charge transfer occurred from P3HT to the polycrystalline Ag foil). The down-shift of the vacuum level of P3HT polymer would result in an increased hole injection barrier ($\Phi_h$) but facilitate electron injection from the Ag Fermi level to the conduction band of P3HT. The hole injection barrier ($\Phi_h$) is given by subtracting band bending (0.2 eV) from the onset of the valence band (0.8 ± 0.05 eV) resulting in a value of 0.6 ± 0.05 eV. By subtracting $\Phi_h$ from the optical bandgap $E_g^{opt}$ (*ca.* 1.7~1.9 eV ) as measured by UV-VIS spectroscopy,[23,44] an approximated electron injection barrier ($\Phi_e$) of 1.1 ~ 1.3 eV is obtained. This makes P3HT a *p*-type material, which is consistence with the previous results on HOPG,[26] ITO,[25] and gold substrates.[19]

IV. CONCLUSION

In this work we have demonstrated that P3HT can be thermally evaporated in vacuum and that its chemical composition and structures are largely conserved as confirmed by FTIR and XPS characterization. The interfacial and electronic properties of P3HT thermally deposited on polycrystalline Ag foils were studied with XPS and UPS using an *in-situ* deposition technique to prevent potential surface contamination. UPS spectra exhibit an increasing vacuum shift with P3HT layer thickness for initial deposition cycles as measured at the secondary-electron cutoff. This shift is caused by the formation of an interface dipole eD and band bending which increases with polymer layer thickness. The band bending taking place at the polymer/metal interface is also manifested in the XPS measurements by the binding energy shift for both C 1s and S 2p peaks. Finally, the localized π-band energy position shifts as a function of P3HT layer thickness to saturate around 2 nm coverage. The value of the hole injection energy is estimated to be 0.6 eV which places P3HT in the *p*-type material category consistent with previous work of P3HT deposited by spin coating and electrospray methods.

**ACKNOWLEDGMENTS**

This work was supported by ARO Grant W911NF-06-1-0295 and by ONR Grant N00014-03-1-0247. We would like to thank Dr. David Cleary for assistance with the FTIR-ATR measurements.

Figures

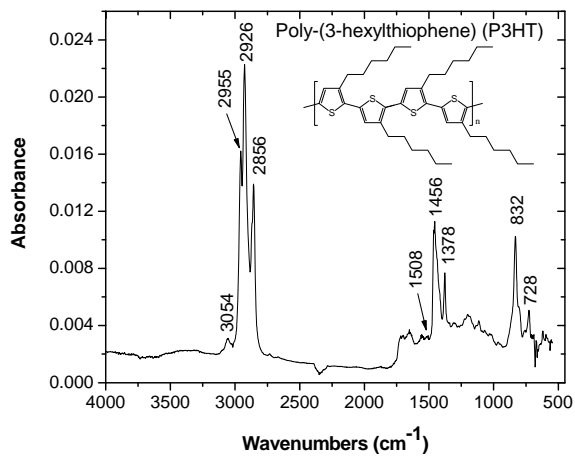 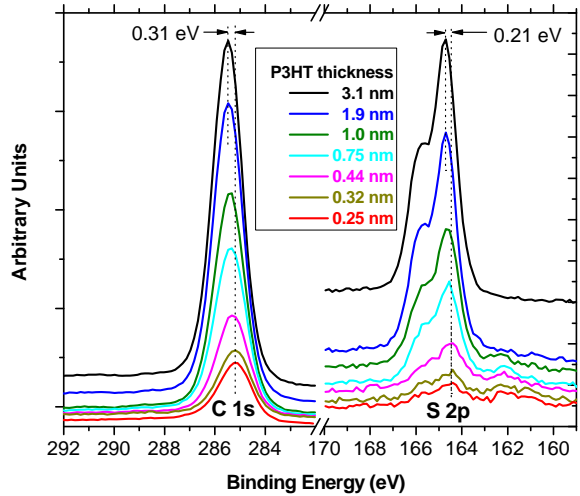

Fig. 1. FTIR spectra of P3HT films on KBr substrates deposited via vacuum thermal evaporation

Fig. 2. C 1s and S 2p photoemission spectra as a function of increasing P3HT thickness on Ag foil

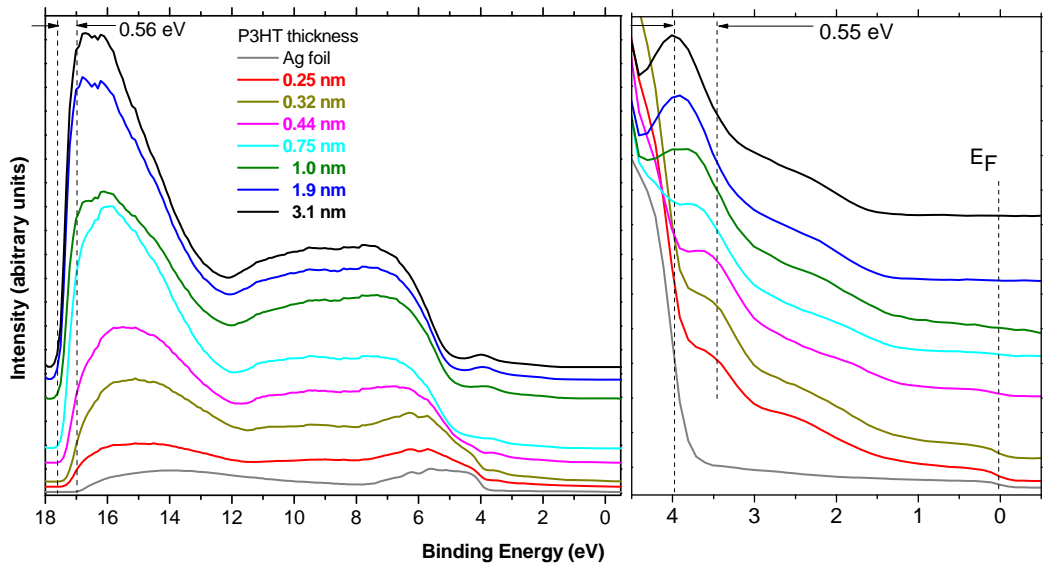

Fig. 3. He I UPS spectra as a function of increasing P3HT thickness. The full spectra as measured in the left graph shows the shift of the secondary cutoff as a function of P3HT layer thickness. The right graph depicts the enlargement of the left plot at lower binding energy region showing the valence band region.

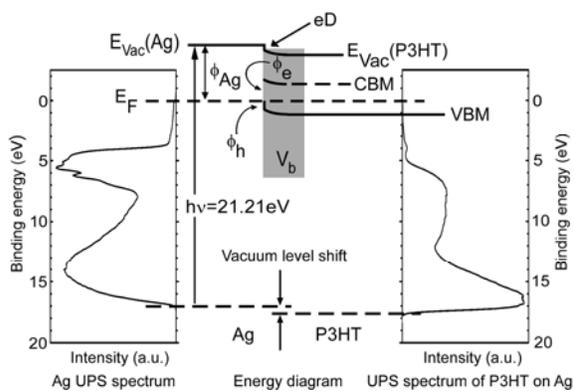

Fig. 4. Schematic energy diagram of the interfacial electronic structure of P3HT/Ag (see details in main text)